\documentstyle[aasms4]{article}

\received{}
\accepted{}
\journalid{000}{}

\begin{document}

{\title{Astronomers and the Science Citation Index, 1981--1997}

\author{David Burstein\altaffilmark{1},}

\altaffiltext{1}{Department of Physics and Astronomy, Box 871504, Arizona
    State University, Tempe, AZ  85287--1504}

\begin{abstract}

The Institute for Science Information (ISI) has generated two lists of
citation information for astronomers that are restricted both as to
the years surveyed for the {\it cited} papers, and the years surveyed
for the {\it citing} papers.  These databases are unique among the
electronically-available citation data in their restrictions of both
citing and cited years.  The main list (P\&A-100) gives citation data
for 62,813 physicists and astronomers whose journal papers were cited
100 times or more from 1981.0 to 1997.5 by papers published during the
same time interval. The second list (AST-top-papers) gives the 200
most-cited papers/year published in refereed astronomical journals
from 1981--1996, as cited in papers in those same journals from 1981.0
to 1998.0.  Astronomer names were selected from those given in the
P\&A-100 list using various sources, including the 1998 AAS Membership
Directory, the 2000 list of the Astronomical Society of India, 
the names from the AST-top-papers list, the list of
astronomers honored by the AAS, National Academy of Sciences and the
Nobel Prize Committee, and the knowledge of this writer.  From this
work an Astronomy Citation Database has been constructed, containing
citation data for 6458+ astronomers.  Various problems, both
substantial and subtle, of producing a reasonably fair citation
database from either the data supplied by the ISI to this author, or
from the Web, are detailed.  Chief among these are whether to assign
either parital or full credit for each author on a given paper.

Whether one is honored with one of the top lifetime-awards given to
astronomers is a strong function of how well your work stands out as
your own.  In particular, we can negatively impact our citation
statistics in two ways.  First, because the ISI does not keep track of
meeting proceedings/books/catalogs, per se, we do not get citation
credit for meeting papers/books/catalogs in the ISI lists. Second, if
we fuzz our identities on the papers, such as publishing papers with 
two or more first initials or through confusion with the names of other
astronomers/physicists.

Name confusion affects this kind of analysis to the extent that that
it would take an enormous effort to disentangle its effects and, even,
then, not all name confusion would be settled.  If sociologists,
science historians and others (ourselves included?) feel that solving
the name confusion problem is worthwhile, perhaps we, as a
professional society, should take appropriate steps.  A ``modest
proposal'' is made that our professional field (and others) go to a
system of uniquely associating an identification number to each author
on each paper.

\end{abstract}


\section{INTRODUCTION}

For the past several decades, we in astronomy have relied on the
Science Citation Index, as compiled and published by the Institute for
Science Information (ISI), as our source for citation statistics of
our papers.  At the dawn of the 21st Century, we in astronomy are
becoming more and more reliant on electronic databases for the papers
we read, and for the citation statistics on those papers.  The two
main internet sites we access for listings of our published papers (as
opposed to preprints) are the Astrophysics Data System (ADS)
(adsabs.harvard.edu) and the Institute for Science Information's (ISI)
Science Citation Index (www.webofscience.com).  

Indeed, the Science Library at our University no longer subscribes to
the ISI's printed Science Citation Index; rather there is now complete
reliance on the ISI's web-accessible Web of Science.  Yet, as this
author has discovered during the research conducted for this paper,
many of us do not clearly understand the contraints and limitations of
either the hard-copy Science Citation Index and the new Web of
Science.

However, what prompted the research done in this paper was the
a new kind of citation analysis produced by the ISI research group
over the past few years.  At least two new citation lists were generated by the
ISI research group before this author contacted them: A list of the
chemists cited 500 times or more and a similar list for
physicists/astronomers.  Each list differs from what we can access
either from the Web of Science or from the hard-copy Science Citation
Index, in that time intevals for both {\it citing} and {\it cited}
papers are specified.  

A French chemist, Dr. Armel Le Bail (Laboratoire des Flouresces, CNRS,
ESA) purchased the ISI's most-cited chemist list (for \$1000), and
posted it on the Web (pcb4122.univ-lemans.fr/cgi-bin/physiciens.pl).
Along with the most-cited chemist list, the ISI sent Dr. Le Bail the
first 1120 names of physicists/astronomers on {\it that} most-cited
list, which Dr. Le Bail scanned into his computer and posted on the
web.  One of the reasons Dr. Le Bail posted these lists on the web was
given by Dr. David Pendlebury of the ISI, who points out that, of the
50 most-cited chemists, 7 have been awarded the Nobel Prize
(cf. Garfield \& Welljam-Dorof 1992).

The interest of the present author in this subject was piqued when
one of his colleagues pointed out that his name was in the most-cited
physicists/astronomers list on the web page of Dr. Le Bail.  Thus
began the journey of this author down the rabbit hole of web-accessible 
and web-generated paper and citation information.  The present paper, 
with its lessons learned and data gathered, is the net result of that 
journey.

Aside from the not-inconsiderable curiosity factor (e.g., most of us
would like to know where we stand relative to others in the number of
times our papers have been cited) and job-related factors (my own
promotion to professor was aided by such a list compiled by a
colleague for one year of citations for our department), why would one
do such a study?  This author can think of several questions one would
like to answer.  First, at the very least, if citation information is
to be used in connection with job-related decisions, should not the
available data be of highest possible quality?  Such data should be
treated like any other data, and investigated as to random and
systematic errors.  Should it not also be clear what assumptions go
into the data being used?  Given that a relationship exists between
being most-cited and winning a Nobel prize among chemists, does the
same relationship exist for honors received by astronomers?  As a
guide to those scientists just entering our field of study, what do
these data tell us about how the way we put our names on our papers, and 
where we publish our papers, influence how we are honored by our peers?

Previous papers which tried to assess citation information for
astronomers (Abt 1981a,b, 1982, 1983, 1984a,b, 1985, 1987a,b, 1988a,b
1989, 1990a,b, 1992a,b, 1996, 1998a,b; Abt \& Zhou 1996; Trimble 1985,
1986a,b, 1988, 1991, 1993a,b, 1996; White 1992; Girard \& Davoust
1997; Davoust \& Schmadel 1987, 1992) were limited by time and data
access to asking statistical questions that are more restricted than
those that now can be addressed electronically.  The methodology
employed by this paper are detailed in Section 2, where we address, in
detail, what one can do, and what cannot do, with the present
databases made available to this author by the ISI as well as those
databases generally available on the web.  The data we have generated
from this study are discussed in Section 3.  The statistical studies
of citations for astronomers are discussed in Section 4, both among
themselves and in comparison to the life-time honors bestowed to
individuals. The main results of this paper are summarized in Section
5, where a ``modest proposal'' is made towards solving the
ever-pervasive name confusion problem.

\section{METHODOLOGY}

\subsection{The ISI Databases: Definition and Restrictions}
%2.1

The first list of most-cited physicists/astronomers generated by
Dr. David Pendlebury and his collaborators and given to Dr. Le Bail,
has a cutoff at 500 citations/name.  When the ``unique-two'' issue was
discovered (cf. Secs. 2.3, 3.1), this writer requested a new list be
generated by the ISI that placed its cutoff at 100 citations/name.  This
new list, provided by Dr. Pendlebury to this writer, gives 62,813
astronomer and physicist ``names'' cited in the ``usual'' refereed
journals at least 100 times or more during the time period January,
1981 (1981.0) through June, 1997 (1997.5), for papers published in the
same journals during the same time period (hereafter referred to as
the ``P\&A-100'' list).  Note that what the ISI provides in its databases
is literally a last name and first initial(s), without uniquely
identifying that name, per se, with an actual individual.

Separately, Dr. Pendelbury's group has generated a list of the top 200
papers cited in astronomy each year from 1981-1996 (``High Impact
Papers in Astronomy, 1981--1996''; hereafter referred to as the
``AST-top-papers'' list) from a set of astronomer-used journals (see
Table~1), for citations made up to 1998.0.  Among the 3,200 papers in
this database are 5,035 astronomer ISI ``names.'' For the purposes of
this paper alone, the ISI has given this author access to these two
databases, and it is from these datasets that this author has compiled
the citation data in this paper for astronomers.  In return, this
author has communicated to the ISI the various issues one finds when one
attempts to do a complete survey of citations for scientists in a
relatively well-defined field of study, such as our own.

Abbreviations for the journals used for the ISI P\&A-100 list are
given in Table~1, divided into four classes (ISI code in parentheses):
space science (SP), condensed matter and applied physics (APP),
optical and acoustic (O/A) and general physics (PHS). Essentially all
of the standard journals in which astronomers publish their papers are
in the SP category. The journal abbreviations come from both the
Astrophysical Data System (ADS) (generally mixed upper and lower case
letters) and from the ISI (an 11-character code, all capital letters).
Those SP journals whose abbrevations are given in bold letters in
Table~1 are used for the AST-top-papers database.  As is evident, the
AST-top-papers list employs the ``usual'' astronomy journals, so the
``names'' in {\it that} database can be all considered as those of
astronomers for this analysis.  In contrast, the P\&A-100 list
includes all of the journals in Table~1, and thus includes far more
physicists than astronomers (as is evident from the marked contrast in
numbers of names in the two lists).

Both the ISI databases have limitations placed upon them in terms of how
the ISI does its business in handling certain well-known ``sticky'' 
citation analysis issues:

First, no meeting papers or books are used for these compilations,
either for the cited papers or the citing papers.  Indeed, if you
currently use the Web of Science's ``Cited Search'' option, you have
to specify at least the first author of {\it any} article for that
search to bear fruit (more discussion on the issues of using this
webpage in Secs. 2.4-2.7).  The ISI is not alone in making apparently
draconian decisions about meeting papers.  The ADS also tends to list
only a few authors on multi-authored meeting papers, as well as for
many journal papers published pre-electronic submission (see
Secs. 2.4-2.7).  It is apparently difficult for electronic databases
to account for authorship of meeting papers/catalogs/books or 
multi-authored papers, in their citation statistics for scientists.

Second, while the authorship of a paper that is cited comes from the
cited paper itself (minimizing spelling errors of authors' names),
the citation for that paper comes from the citing paper.  Human error
being as it is, citations in the citing papers can sometimes be in 
error.  This leads to what is, in effect, a random error for the
citations that is proportional to the number of citations for a given 
paper (i.e., more hands in the pot, more likely an error).  An
estimate of this random error will be made in Sec. 2.6.

Third, only the first 16 authors of a given journal paper are used in 
the two ISI databases used here. This decision was made by the ISI in the
interests of economy, and from the fact that relatively few 
physics/astronomy papers have many authors.  Fortunately, one can use
the AST-top-papers list to make corrections to the P\&A-100 list to
account for this exclusion (Sec. 2.2) so that the resulting error
in citation statistics is neglgible.

Fourth, the long-standing decision of the ISI is that every author of
a paper be given full credit for that paper.  Whether or not one
agrees with this decision, this is what you get when you
access the Web of Science or the ADS for citation purposes. Whether or
not this author (or a reader of this paper) agrees with this decision
by ISI or not, this is what the ISI does to generate the Web of
Science and the data lists used in the present analysis.  It is also
in keeping with the philosophy they used for the hard-copy Science
Citation Index, but could not completely employ in that presentation
mode.  This important issue is discussed in detail in Sec. 2.4.

Fifth, the lists are not lists of individual astronomers and
physicists, but, as stated above, the ``names'' as given on their
papers.  These ``names'' uniformly have a last name and first
initial(s) only.  In cross-correlating these two datasets with lists
of actual astronomers, name confusion is inevitable if all we have for
identification are the ISI names. Many people share the same last
names and first initial(s).

\subsection{A Step-by-Step Name Search Process}
%2.2

The two ISI databases provide citation data for astronomers in a
complementary way.  Comparison of the names of astronomers in the
AST-top-papers list with those in the P\&A-100 list shows that not all
astronomers who have 100 or more citations are included in the
P\&A-100 list, owing to the multi-author problem (``sticky'' issue 3,
above). Comparison of the names of astronomers in the AST-top-papers
list to those in the AAS 1998 Membership Directory finds many
astronomers not in the AAS Directory.  Hence, by using the P\&A-100
list as the primary data source, and the AST-top-papers list as the
secondary data source, we can assemble a more complete citation
database for astronomers than by using either ISI database alone.  The
assembly of the citation database for astronomers was done through
five time-consuming steps, using all of the information, both written
and electronic, available to this author.

Step 1. The names of astronomers in the 1998 AAS Membership Directory
were compared with the last names, first initial(s) in the P\&A-100
list. Those in common were noted.  The AAS Directory was used as this
is the only membership directory for astronomers generally available.
Subsequent to placing a copy of this paper on the Web, this author
did the same comparison for those astronomers in the 2000 directory
of the Astronomical Society of India.

Step 2. For each of the found ``names,'' a web search was made using
the ADS to see whether more than one person, physicist or astronomer,
could have that ``name.''  The ADS was used for this search, as it
gives, in many cases, first names for individuals.  Depending on what
was found on the ADS, the listed ``name'' was either uniquely assigned
to an astronomer, or that ``name'' was marked as confused with other
astronomers/physicists.  At this point in the process, a preliminary
Astonomy Citation Database (ACD) was assembled ($\sim 3700$ names).

Step 3. To further help to overcome the obvious AAS-driven, North American bias
in the preliminary ACD, the ``names'' of astronomers in the
AST-top-paper list were compared to the names in both the P\&A-100
list and in the preliminary ACD.  Those AST-top-papers ``names'' found
in the P\&A-100 list but not yet in the ACD were added to the ACD via
the same Web search process as in Step 2 above ($\sim 2500$ names).

Step 4. To overcome the 16 author/paper limit of the ISI databases, a
search was made of all papers (105) in the AST-top-papers list that
had more than 16 authors.  These included papers with as few as 15
citations and as many as 908 citations.  The remaining authors on
these papers were then found either by looking up the paper on the Web
of Science or from the actual hardcopy of the journal.  As before, the
additional data for astronomers was added into the ACD via the process
outlined in Step 2.  The citation data for new astronomers ($\sim
170$) found by this process, and the citation data that was added to
astronomers already on the ACD, are noted as ``A16'' in the ACD.

Step 5. We generated a list of astronomers (278) honored by the AAS
since 1949, honored by the Nobel Committee and/or who are current members
of the U.S. National Academy of Sciences.  Any astronomer in this list
not already in the existing ACD ($\sim 50$) was added to that list if the
astronomer's name was also in the P\&A-100 list.

The citation data from the two ISI databases were modified in just one
way to handle the special cases of three well-used data catalogs that
were published during the cited time period (1981.0 to 1997.5): the
Third Reference Catalog of Bright Galaxies (de Vaucouleurs et
al. 1991), the Revised Shapley Ames Catalog (Sandage \& Tammann 1981,
1987) and the Bright Star Catalog (Hoffleit 1982).  The printed
Science Citation Index and the Web of Science were both used to
estimate ($\pm 100$) the number of citations to add for these catalogs
for the individuals involved: 1000 citations for the Third Reference
Catalog, 1250 citations for the Revised Shapley--Ames Catalogs, and
1395 citations for the Bright Star Catalog.  Citation data for these
catalogs follow the same rules as for the citation data in the ISI
lists --- only the those citations as given in papers in refereed
journals during the stated citing time period.

\subsection{Name Search Leads to Name Confusion}
%2.3

The ISI procedure of taking the names of the authors {\it from the
papers cited}, rather than from the citing papers eliminates most, but
not quite all, spelling errors.  Of the few specific instances of
misspelling caught by this author in the two ISI databases, most have
been fixed by the ISI.  There still remain three individuals whose names
that are likely mispelled in the journal proceedings (or mis-scanned
by the ISI): RP Kirshner/Kirschner; A Dyachkov/Dyatchkov; H Luhr/Luehr.
The case of RP Kirshner is curious, as in 1998, the ADS recognized
that both spellings are associated with the same person. As such,
while citations are added together for Kirshner/Kirschner, the numbers
of papers cited is kept at the value listed for Kirshner.  In the
cases of A Dyachkov and H Luhr, the mispellings are quite evident, so
the papers and citations for both spellings are added together.

The same person can also have different last names, owing to marraige.
Four persons having two names in the databases were so identified (with
kind aid from V. Trimble): J Bland--Hawthorn/Bland, MJ Rieke/Lebofsky,
S Viegas-Aldrovandi/Aldrovandi and S Collin-Souffrin/Collin. The
citation data for the first three of these astronomers were added
together, as the different names are on different papers.  The
citation data for S Collin-Souffrin cannot be added together with
those for S Collin, as the S Collin name is confused with that of another
astronomer as well as with names of physicists.

In an attempt to disentangle as much of the name confusion as
possible, this writer took advantage of an aspect of the ADS website
that is apparently not yet available on the Web of Science website:
The ADS website will search all names using the string of letters it
is given, and will do so for an exact name match even if you do not
give the whole name.  Moreover, the ADS will often give the first
names of the individuals found.  As such, it was possible to search
both the astronomy and the physics ADS websites for each ISI ``name''
suspected to be name-confused.  This search was eventually done for
almost all ``names'' thought to be those of an astronomer.  When an
unambiguous first name is found for a last name having only one first
initial, that first name is noted in the ACD.  In addition, all last
names that are in fairly common usage were searched for name confusion
for all sets of first initials.  Some name confusion was solved by
comparing citation data for astronomers as obtained from the P\&A-100
list with those obtained for the same astronomers in the
AST-top-papers list. In the former list, astronmers and physicists can
have their names confused; such is not the case in the latter list.

It was quickly found in this analysis that some astronomers are listed
under two or more sets of initials in the ISI databases (e.g. JP
Huchra also commonly has J Huchra on his papers).  In the 357 cases
currently in the ACD for which this writer could identify the author's
last name with two or more sets of first initials, {\it and} that set
of initials/last name are all uniquely identified with an individual,
the citations are summed for all sets of initials with the same last
name.  The number of papers is summed as well, as the ISI would find
these as separate papers from those with another initial for the given
last name. 

Name confusion can occur for two reasons.  First, many last names are
common in each culture (e.g., Smith, Jones, Wang, Suzuki, Singh), so that
those individuals with only one, or even two, first initials can have
their names confused with those of other astronomers and, especially
physicists.  In certain cases, astronomers related to each other may
have the same last name and first initials.  Second, given that we
find many astronomers using two sets of initials for their papers, we
have to allow that other astronomers may do this, but that one or both
sets of names they use can be confused with those of others (who also
may or may not use two sets of initials).  This then leads to all
possible combinations of confused/unconfused names: one set of
initials for a given last name uniquely identified with an individual,
the other set confused; both sets of initials confused; a last name
with one first initial having possible confusion with 2 or more (up to
6) names with that first initial plus different middle initials.  In
other words, far more than 357 astronomers publish using two or more
sets of initials for their papers. 

If this all sounds confusing, well, it is!

\subsection{Methodology Summary}
%2.4

Investigating the citation data for astronomers is, indeed, a trip
down the rabbit hole.  To help make sense of this trip, here we answer
seven questions one can ask about this kind of citation survey:

1. Which journals/meeting--proceedings/books does one choose from
which to gather citations? The ISI has defined the answer. Only
refereed journals as defined by the ISI (Table~1) are used here, both
for papers cited {\it and} for papers doing the citations.  Thus,
papers published in meeting proceedings, and authorship of books and
catalogs, are not used for citation statistics (Sec. 2.2).

2. The papers that do the citations are published during what time
periods?  The ISI defines these papers to be published between 1981.0
and 1997.5 in all of the journals in Table~1, including both physics
and astronomy journals.  The AST-top-papers list covers the citing
papers during the period 1981.0 through 1998.0 (i.e., 1/2 year longer
than the other list), using the astronomy journals whose abbreviations
are in bold face in Table~1.

3. Papers during what time period are cited?  The ISI defines this
time period for each list slightly differently.  For the P\&A-100
list, it is from 1981.0 to 1997.5, same as the paper-citing time
interval.  For the AST-top-papers list, it is for papers published
from 1981.0 to 1997.0, ending one year before the paper-citing period.
In both cases, the cited papers are from the same journals as those
that do the citing for each database. Again, it is stressed that
meeting proceeding papers are neither cited nor are citations taken
from them by the ISI.  Hence, no meeting proceeding data are included
in ACD.

4. Credit only the first author, or credit all authors on a given
paper?  The ISI standard practice for its Web of Science, and for its
generation of citation lists is that all authors on a journal paper
are credited for each paper.  As stated earlier, this is {\it not} a
change in ISI policy, but rather full expression the ISI policies
given the freedom of the web.

5. Fractional, or unitary credit of citations for authors on
multi-authored papers? The ISI gives each author of a journal paper 
full credit for that paper (i.e., unitary credit), up to 16 authors
per paper.

6. How does one handle name confusion when only first initials and
last names are available?  This is handled in a complicated manner,
using all available means available.  The sobering fact is that even
having full first names available does not completely remove the
confusion issue, as many individuals also have the same first and last
names.

7. What is the accuracy of the citation estimates?  As dicussed below
(Sec. 2.6), the random error in the ACD citations is proportional to
the number of citations received, at about the 4\% level.  Other
small, systematic effects exist as well, stemming from several
different issues.

Of all of the choices that the ISI makes for its databases, the
assignment of one paper credit to each author of a paper is the
most-controversial.  While this writer agrees with this choice, 
others with whom the author has discussed this paper do not.
The plain fact is that any present or future study that uses the ISI
citation database is using citation data which assigns each author of
a paper all of the citations for that paper.

Given the lively controversy this decision by ISI has engendered among
this authors colleagues, this writer feels it necessary to state why
he agrees with ISI's choice in this matter. First, and foremost, this
writer does not know of a universally-accepted fair way to give
fractional credit.  On the small sampling of astronomer colleagues,
whether or not one votes for fractional credit seems to depend on
whether or not one has been involved in a multi-authored paper.  Those
of us who have been involved in multi-astronomer projects tend to vote
for giving full credit to each author of a paper.  Those who have not
been involved much with such projects tend to vote to give fractional
credit.

Yet, if ISI, or any reader of this paper, decided to give fractional
credit for the citations of a paper, how would such fractional credit
be calculated?  Would it be fairer to divide the number of citations
strictly by numbers of authors, and give each author fractional
credit? Or do we try to credit some authors (say, first author) more
than others?  As shown in Sec. 2.6, if we take the most
straightforward way to calculate citations for individuals in a
fractional sense --- even division by numbers of authors --- there
would be substantial revision of names among the top-cited astronomer list.

\subsection{Two ways we lessen the impact of our papers}
%2.5

In the compilation of the astronomer citation list, this author
has found that getting credit for the papers you publish can be 
lessened in two ways.

Both the ISI databases and the ADS (Sec. 2.1) have problems in
registering citations for papers published in meeting proceedings. 
While books and catalogs can be found in the Web of Science via a 
``cited reference'' search of the correct first author name and year
of publication, these citations come only from those made in journal 
papers.  So, while meeting papers are cited there, these data are not 
incorporated into the main ISI databases.  Meeting papers do not
figure at all into the citations statistics for the Web of Science, 
nor for the two ISI databases given this author to develop the ACD.

The other way credit for citations of your paper can be harmed is if
your name is either confused, or if you use more than one first
initial on your papers.  For those of us with common last names,
especially our Asian colleagues, solving name confusion will not easy
(cf. Sec. 5). If you permit your papers to be published with two or
more sets of first initials, the Web of Science will put you in two or
more different places, fuzzing the credit you will get for your
citations.  This issue was one thing when we used the Science Citation
Index hardcopy, as we could easily see the two different entries.
Such is not the case when you access the same data electronically.
Indeed, if using electronic databases, finding citations data for an
individual who uses different first initials requires knowledge
aforethought that this problem exists.  While someone working in the
field of study can sort this problem out with a lot of effort, it is
impossible for people {\it not} working in that field of study (such
as the ISI personnel) to do the correct sorting.  

\subsection{Citation Errors: ``Random'' and Systematic}
%2.6

A ``random error'' infects the ISI databases owing to the manner in
which the ISI gets its citations per paper: the journal volume, page
number of the cited papers are taken from the citing papers.  Human
error being what it is, a certain percentage of those cites give the
wrong journal number, wrong page number (sometime switching one for
the other), confuse ApJ with ApJL or ApJS, etc. One does not discover
that this can lead to errors in the ISI databases until one accesses
the Web of Science, pushes on the general search button, then opt for
a ``cited ref search.''

As opposed to the ``general search'' mode, ``cited reference search'' 
is best used by entering both the first author and the correct year
of publication.  What one then finds is a list of the citations
for the papers of that author. Those that are underlined in blue
you will also find in the general search.  Those not highlighted
either are misentered in a citing journal paper, or are referenced in
citing journal papers to non-journal papers, private communications, 
books, catalogs, etc.  In cited reference search mode you will see all 
of the misentered entries for a given cited paper.  

This writer has accessed the cited reference search for 36 papers in
Table~3 to estimate the misentry incidence as a function of journal.
Note that statistics cannot be done for the citation data in the ACD
per se, but rather for the current (2000.0) citation data for the
papers involved.  Citation data for individual papers is affected as
much as 126 misentries for 989 1999.9 citations (for 1992 ApJ Letter
paper of Smoot et al.) to a low of 2 misentries for 532 1999.9
citations (the A\&A paper of Renzini \& Voli 1981).  If we express
these errors in terms of percentages, they range from a high of 19\%
to a low of 0.004\%, with a mean of 4\%.  This error is strongly
journal--dependent, being the most for the ApJL (with a relatively
high percentage of papers not citing the journal as ApJL), and lowest
for Nature and PASP.

Given that none of us publish our papers solely in one journal, a
reasonable estimate for a one-sigma random error of citations in the
ACD is 4\%.  While this random error can be corrected in principle, in
practice it would be highly labor-intensive, requiring a scientist
from our field to work directly with the ISI to make the corrections,
paper-by-paper.  As such, a one-sigma random error of 4\% in the
number of our citations can be viewed as the dues we pay for using the
digital computers to do the citation calculations.

Several systematic errors also affect the citations in the ACD:
 
Omission of astronomers: Non-inclusion of astronomers at the low end
of the ACD comes about in a somewhat convoluted manner. 173 astronomer
names from the ``A16'' analysis were excluded from the ACD because the
ISI names of those astronomers had fewer than 100 citations accredited
to them in the A16-corrected AST-top-papers list.  The AST-top-papers
list references only a subset of the journals used for the P\&A-100
list, of which only a subset of {\it those} papers are used.  Hence,
it is likely that the names of at least some of these astronomers
would be in the ACD if all papers on which they are authors had been
included in the P\&A-100 accounting (i.e., their other papers could
total up to 99 citations).  As such, these 173 astronomer names are
given in a separate ``honorable mention'' list that will be supplied
electronically.

Separately, inclusion of people who became deceased during the sampled
period is handled both from the memory of this writer, and through the
comparison of the astronomer names in the AST-top-papers list to those
in the P\&A-100 list.  It is not expected that many such astronomer
have been omitted from the ACD.  As an additional check, a scan of the
full P\&A-100 list by this writer of those ISI names with 3000 or more
citations uncovered no other astronomer names than those given in the
ACD.

Wrong papers: One can compare the authors on the papers in Table~3 to
the astronomer names in Table~2, to see that papers that publish wrong
results do not substantially contribute to the citations for these
astronomers.  Moreover, one can verify that very few of the 3000+
astronomers in the top half of the ACD have reputations built on the
publication of wrong papers.  The conclusion of this paper is that
citation of wrong papers negligibly influences the citation
statistics.

Self-citation: The present available websites make it difficult to
quantiatively assess the effect of self-citations on the citations for
a given astronomer.  What is important here is the variance around the
percentage of our citations that are our own papers, as this is
something we all do.  On can most directly assess this for individual
papers, which this writer has done for about a dozen, many-cited
papers.  The result of this non-statistical sampling is these papers
self-cite in the range 3--15\%, with a mean about 8\%.  Since it is
variance about this mean that affects the citations of one astronomer
vs.  another, we can expect a variance of 5--7\% in citations that can
be ascribed to self-citation.  Whether or not to assign a
``self-citation'' variance of $\sim 6\%$ to the ACD data (to then be
added in quadrature to the random error estimate of 4\%) is a matter
of choice.  As with giving/not giving full citation credit for each
author of a paper, honest people can differ on whether or not to
account for self-citation variance in the ACD.

Fractional or Full Credit: While a scientifically correct test of this
issue cannot be made from either the Web of Science (no restriction on
citing years) or from the two ISI databases.  As such, compiling a
list that is complete as the ACD that gives only fractional author
credit is impossible without the full cooperation of the
ISI. Nonetheless, an illustrative differential test can be made using
the data in the AST-top-papers list.  This writer took the top 13
individuals cited in the AST-top-papers list (which are {\it not
necessarily} the same as the top-cited authors in the ACD), and
counted the number of authors of each paper for each individual,
as well as counted the number of papers that individual was first
author, and the number of papers the name of that individual was
placed in alphabetical order (for papers with 3 or more authors).  To
this list this writer added the same data for three other astronomers known
to this writer to mainly publish significant, single-authored papers.  

For a first test, fractional credit was given by dividing the number
of citations for each paper by the number of authors on that paper,
and then summing the fractional credit for each individual.  The ratio
of (fractional citations/full citations) so obtained ranged from 0.05
(for an individual publishing mainly with large groups) to 1.00 (for
an individual with 2 sole-authored papers), with a median of 0.30.  If
these 15 individuals were the only ones in the database, a substantial
reordering would be done from the full citation credit list.  The net
result would be to keep 7 of the 13 top-cited individuals in the top
positions, but move 6 individuals from lower on the full credit list
to near the top of the fractional credit list.  

For a second test, we find the ratio of papers for which the author is
first or sole, to the number of individual's papers included in the
AST-top-papers list.  This ratio varies from a high of 1.00 to a low
of 0.067, with a median of 0.27.  If we then would calculate
fractional credit by giving more credit to the first author of a
paper, the order of listing of individuals would be different from
that of either full credit or strict fractional credit lists.

The bottom line here is that there is simply no way one can calculate
the number of citations for individuals that everyone can agree with.
Depending on which way one chooses, the result will be different.
Rather, given the now wide-spread use of the Web of Science, it seems
most logical to accept the practices of the ISI in this regard in
giving full citation credit to each author on a given paper.

Interestingly, how the ISI has presented its database has been a
product of techonology.  From the start, for the the printed Science
Citation Index, ISI obviously had to make hard choices of how to
present data, both in number of letters for names and number of
authors per paper used.  One choice that was made for the printed
version was to list papers under only first authors in the Author part
of the Index.  All authors of a paper (up to an author limit given in
the Index explanations of that year/summary list) would then be
referenced to the each paper in the Source part of the Index.  As
such, previous investigations of citation-related issues would,
indeed, have just found first-authored papers if only the Author part
of the Index was used.

In contrast, the Web of Science now permits the ISI to show all of its
data for each author.  This means that when one accesses a given
author name, one gets the full citations for each paper that author is
on, not just the ones for which that author is first.  The ISI notes
such other papers by preferencing the Web-cited reference with several
dots.

In summary, errors in data lists are introduced due to human error or
human choice, whether using a computer or not.  When data lists are of
numbers, errors or choices are one thing; when data lists are of the
accomplishments of real people, errors or choices are quite another
thing.  It is likely that human errors of omission and comission in
ACD exist.  Unforunately, there is no easy way for this writer to find
all of them.  The most serious of the likely systematic errors in the
ACD concern exclusion of astronomers from the list, while the random
errors of citations in the list are at about the 4\% level (modulo
accounting/non-accounting for 6\% self-citation variance).  The most
intractable issue is to whether or not give full citation credit to
each author.  Whether or not one views this as an error in the data
base depends on ones opinion.  Interestingly, if one had only used the
Author section of the hard-copy Science Citation Index in past years,
one would have only found those papers for which the author was first.

While the current version of the ACD has been constructed to be as
complete as possible, it is the ongoing aim of this writer to make
this list as complete and accurate as is reasonably possible.  Given
that the time periods of citing and cited papers are fixed, this is
doable. Towards this end, the reader is encouraged to contact this
writer if the reader feels a name should be in the ACD that is not, of
if one feels the data entered for a person is badly in error.  All
such cases will be addressed individually by this author and
corrections to the ACD made, if warranted.

\subsection{Are the ISI Lists the Best We Can Do?}
%2.7

The ISI lists do something important that this writer has not yet
found among the available websites or in written form: Specify both a
time interval for {\it when} the papers are published and a time
interval for {\it when} the papers citing those papers are published,
and sum all of the found citations per author name.  This permits
``snapshots'' in time of citations to be assembled, of which the
current ACD is just the first.

The Web of Science gives the number of citations per paper, but not
per author for that year.  This latter number is something one would
have to manually extract from the information given.  The Web of
Science also does not currently permit the citing years to be
restricted.  As such, when checking for the random error problem
(Sec. 2.4), one could only assess the errant citations against
ISI-registered citations at the current time.  However, a specific
interval for the years the papers of an astronmer were published {\it can}
be specified.  Hence, what one gets is a running number of all
citations of a given paper up until the date you access the citation
information.  In absence of a world-wide directory for astronomers, It
is essentially impossible to predefine all astronomer names to do the
kind of analysis with the Web of Science that one can do with the 
P\&A-100 and AST-top-papers databases.

The ADS website has a button one can push which will give the total
citations per astronomer from their database (such a button is not
available for the physicist webpage, though).  By pushing this button
you get a summary total of citations for that astronomer, and a list
of the citing papers, but not the number of citations per paper, nor
the number of cited papers, nor a list of cited journals/meetings/etc.
A perusal of citing papers for several individuals indicates that
papers in meeting proceedings are only included in the ADS list if
those proceedings could be accessed electronically.  Combine this with
the problems the ADS has in listing authors for multi-authored,
pre-electronic submission papers, the citation information one
gathers from the ADS website has more systematic problems within it
than that we can get from the ISI website.  As with the current ISI
website, the current ADS does not permit defining a specific range of
years for citing papers.

Comparing numbers of citations for astronomers between the P\&A-100
and AST-top-papers lists shows that the P\&A-100 list always has more
citations, with comparable citations for the most-cited individuals.
This indicates that the half-year sampling advantage of the
AST-top-papers list is more than compensated by the greater
completeness of the P\&A-100 list in terms of journals covered.
Suprisingly, it is also found that the ADS citations are generally
less than those of the P\&A-100 list for the same astronomer (at least
in late 1999).  This difference could be a result of referencing only
a restricted number of journals (the ADS currently accesses only
astronomy journals for astronomers, as does the AST-top-papers list).

\section{THE DATA PRODUCTS}

\subsection{The Astronomy Citation Database (ACD)}

The Astronomy Citation Database (ACD) is comprised of the names of
actual astronomers that are associated with the ``names'' in the
P\&A-100 and AST-top-papers.  For each associated astronomer name we
give the ``name'' as given in the ISI databases: last name used (ISI
format without hyphens), first initial(s); the total number of
citations received, the total number of papers cited, and the average
citation/paper (cite/pap).  Each author of a paper is given full
credit for the citations of that paper.  First names are given for those
astronomers for which first names are known/could be found.  The
citation data are adjusted for the ``A16'' issue, the three specified
data catalogs (Sec. 2.2), the changed/mispelled names, and for those
individuals who use two or more first initials.  All additions to the
citation data from the P\&A-100 list are noted for the P\&A-100
``name'' commonly used for that astronomer.  Note that when the
P\&A-100 citation data has been modified for a given ``name,'' the
added data are specified as well.  If one wishes to see the original
P\&A-100 data for a given person, subtract the number of citations and
number of papers that were added. 

For statistical purposes we divide the ACD into three subsets, owing
to the degree to which each ISI ``name'' can be associated with a
unique astronomer:

``Unique-One:'' These are the astronmers whose last names {\it and}
first initial(s) are unambiguosly associated with a single,
identifiable person.  These are most often authors whose names have
two or more first initials and those with uncommon last names.  The
term ``unique'' here refers to the fact that a unique person is
identified with a unique last name and first set of initial(s).  There
are 4617 ``unique-one'' astronomers in the ACD.

``Unique-Two:'' These are the astronomers whose last names are
unambiguously associated with two or more sets of first initials.
Again, the use of the word ``unique'' here denotes association with a
unique, identifiable person. While there are 357 ``unique-two''
astronomers the ACD, the number of astronomers using two or more first 
initials for their papers is likely much more.

``Confused-Named:'' These are the 1484 cases for which unique
ownership of a given last name and first initial(s) is intrinsically
confused.  Here the term ``name'' literally means a ``name,'' not a
unique person.  As stated earlier, name confusion occurs both because
a particular last name is confused for the first initial(s) used, or
because one set of initials for an individual is confused, but another
set is not (the other side of the ``unique-two'' issue).  Of the names
that are confused among astronomers and/or physicists, 398 combine one
version of the name confused, and the other version uniquely-identified. 
Hence the same person might be found in two places in the ACD.  In addition,
there are 94 others pairs of names that have the same last name, two sets
of first initials, and both sets of initials/names are confused among
those of astronomers and/or physicists.  Finally, there are 40 sets in
which one name with one first initial can combine with 2 or more names
with the same first initial but different middle initials, and {\it all}
sets of this name are confused among astronomers and/or physicists.
Hence, nearly half of the confused-named cases in the ACD could involve
confusion among ``unique-two''--related issues.  (Again, sorting
through confused names is confusing!)

To the main ACD we add an ``honorable mention'' list comprised of the
available AST-top-papers citation data for 173 astronomer names from
the ``A16'' analysis whose names were not found in the P\&A-100 list
(cf. Sec. 2.6).

The full ACD is provided in electronic form from this writer, and will be
made available through the Astrophysics Data Center.  For the purposes
of further statistics, in Table~2 we publish those unique-named and
confused-name astronomers cited 3000 times or more.  It is possible to
separate out those names least-confused for top-cited astronomers in
two ways.  First, compare the number of citations between the P\&A-100
list and the AST-top-papers list.  Second, in the case of
physicist-related name confusion, use the ADS and the P\&A-100 list to
check for citations of co-authors on physics papers.  These tests have
only been applied for those ``names'' with 3000 or more citations.
Astronomers with confused names who have a large ratio of
AST-top-papers citations to P\&A-100 citations, and are not found to
have much physics confusion, are designated as ``Cg.''  The Cg
designation indicates that while the astronomer's name is confused, a
substantial number of citations, if not the vast majority, are for the
identified astronomer.

In addition, there are also several possible persons that {\it could}
be among those with 3000 or more citations, but are not owing to name
confusion.  These include B/BA Brown, J/JH Lee, J/JW Lee, J/JM Stone,
J/JC Wang and R/RW Wilson.  The multi-faceted properties of name
confusion also affects the number of citations for
individuals/confused names already in Table 2, with one confused
version of the name qualifying for the list, while the other does not:
CL Bennett has C Bennett with 108 citations, M Cohen has MH Cohen with
1862 citations, RF Green has R Green with 740 citations, RH Koch has R
Koch with 1494 citations, and DW Murphy has D Murphy with 110 citations.

The data in Table~2 are divided by ACD subset, in descending order of
number of citations/name: Column (1) -- the last name of the
astronomer, no spaces/hyphens, in the ISI standard format for last
names; (2) -- the first initial(s) of that name (most commonly cited
in the case of names in the ``unique-two'' subset); (3) -- the
citation/paper ratio (cite/pap); (4) -- the number of papers cited,
including all papers that can be attributed to the individual or name;
(5) -- the total number of citations for these papers; (6) -- Typ is
the sample into which author is placed (1 = ``unique-one,'' 2 =
``unique-two,'' C = ``confused-named'', and Cg as explained above);
(7) -- life-time honor awards: Nobel prize, National Academy
membership, Russell prize, Heineman Prize, with the last two digits of
the year awarded given with the code; (8) -- country of citizenship,
other than U.S.A., coded as given in Section~4; (9) -- comments,
including: for astronomers in the ``unique-two'' subset, the other
initial(s) used by the author (including mispellings), and for two
names (N/NZ Scoville and R/RS Ellis, the fact they were put into the
Unique-2 category is based on comparison of P\&A-100 to AST-top-papers
citations); for astronomers in the ``unique-one'' subset, first names
and middle initials (if any) for those astronomers using just one
first initial, papers added due to mispelling of last name (+sp:);
added catalog citations (+ca:); and corrections for the ``A16'' flaw
in the original ISI databases.  The ``confused-named'' subsets give
the first names of the astronomers having this/these initial(s) (if
the names easily fit into the table), as well as an indication (``+
others'') if there are physicists with the same last name and set of
first initial(s).

\subsection{Top Ten Astronomy Papers Cited, 1981--1996}

While the full list of 3200 papers in the AST-top-papers list is an
item the ISI has made available for purchase only (\$1995), the ISI
has given this author permission to print the top 10 papers cited for
each year.  Table~3 does so for the years 1981--1996 in standard
reference format: Column (1) given the number of citations for that
paper; (2) the absolute ranking of the paper in the full list of 3200
top--200 cited papers of each year sampled (if number of
citations/paper is 90 or over, this is an absolute ranking; if less
than this number, no absolute ranking is given, as the AST-top-papers
list is not complete for papers with 89 citations or less); (3) the
ranking within a given year; (4) the standard astronomical reference
for the paper, using the ``et al.''  for other authors when the number
of authors exceeds five (NCC is NUOV CIM C = Nuovo Cimento della
Soc. Italiana di Fisica C, Geophysics); (5) the year of publication;
(6) the journal in which the paper was published; (7) the journal
volume; and (8) the starting page number.

\subsection{The Honored Few}

We in the U.S.A. astronomical community have developed a large number
of ways to honor those among us who we feel have done outstanding work
in their scientific fields.  These awards are separate from that
considered the ultimate of our profession --- the Nobel Prize.  As
such, we can go further in our comparison of citations to honors than
can likely be done in other fields (cf. the Nobel prize comparison for
chemist made by Garfield \& Welljams-Dorof 1992).  While honors are
not awarded solely on the basis of numbers of citations, it is
certainly the case that the more our papers are cited, the more likely
our work is known to others.  As such, one expects a good correlation
between the names of those astronomers who have been honored and those
most cited.  The ACD was partly assembled to test this hypothesis.

All nine astrophysicists to whom the Physics Nobel Prize has been
awarded since 1964 are included among the 6458 names in the ACD: CH
Townes (1964); HA Bethe (1967): A Hewish (1974); AA Penzias, RW Wilson
(1978); S Chandrasekhar, WA Fowler (1983); JH Taylor, RA Hulse
(1993). That their Nobel prizes do not correspond to their positions
in the ACD owes much more to the fact that for all but the 1993 Nobel
prize, the work for which the prize was given well pre-dates the time
interval sampled here.  Additionally, name confusion affects the
citations for JH Taylor and RW Wilson, both of whose names are found
in the ``confused-named'' subset of the ACD, illustrating well {\it
that} problem.

A total of 278 astrophysicists have been honored either by the Nobel
Prize committee (since 1964), and/or the American Astronomical Society
and its scientific divisions (since 1949), and/or by current
membership in the National Academy of Sciences (in either the
Astronomy, Physics or Geophysics sections).  While only the lifetime
awards are given in Table~2 for individuals so-honored by them, the
electronically-available ACD lists all of the awards given by the AAS
and its divisions.

Of the 278 individuals honored by the AAS, NAS memberships or Nobel
Prizes, 41 are not in the ACD.  Of these 41 individuals, 29 received
their last honor before 1980, hence most of their papers are likely
published before 1981.  Three other honored astronomers not in the ACD
received only one honor that is more related to public service than to
science, per se.  Of the remaining nine honored astronomers not in the
ACD, it is likely that most of their distinguished work was published
before 1981.

\section{The Statistics of Citations}

\subsection{Unique-One vs. Unique-Two vs. Confused-Named}

Three data products are produced from this analysis: the number of
citations per astronomer or astronomer ``name;'' the number of papers
cited for these citations; and the ratio of the number of citations to
number of papers (cite/pap).  Figures~1a,b,c show the histogram
distribution for the unique-one, unique-two and confused-name lists
for each of these parameters. The data that go into these figures are
given in Table~4.

The adopted procedure of handling the number of papers for the
unique-two astronomers is verified by the fact that the unique-two
authors have a satistically {\it higher} cite/pap ratio than do the
unique-one authors.  This confirms the choice of adding together both
papers and citations for unique-two astronomers. In contrast, it is to
be expected that the confused-named subset have a signficantly lower
cite/pap ratio than the subset of unique-one astronomers.

As shown in Figure~1b, the distribution of the number of citations is
nearly identical for the astronomers in the unique-one and
confused-named subsets, but skews to signficantly higher values for
those in the unique-two subsett. In Figure~1c we see that the numbers
of papers cited have the lowest distribution for unique-one
astronomers, but are of comparable higher values for the
confused-named and unique-two astronomers (note the relatively large
number of confused names with 300 or more papers).

Confused-named astronomers have a systematically lower cite/pap ratio
than either of the unique-listed astronomers, while having a similar
distribution of numbers of citations as those in the unique-one list.
This difference was noted early in the analysis and was used to help
seek out confused names by investigating the ownership of those names
with low cite/pap ratios.  The most cited names in the ACD are those
astronomers with asian names whose ISI names are multiply-confused,
even among astronomers.

The big surprise of this analysis is the finding that the statistics
for the unique-two astronomers exceed those for the unique-one
astronomers in {\it all} three catagories.  This difference is also
seen when median values for these parameters are compared: The median
values for number of citations are 381, 425 and 863 for unique-one,
confused-named, and unique-two astronomers, respectively.  Similarly,
the median values for papers cited are 24, 40 and 38, and for cite/pap
are 18.3, 11.5 and 24.2. 

Part of the difference in the citation pattern for unique-one authors
vs. unique-two authors is consistent with the idea that the median
value of citations is likely be twice for people cited in two places
in the original list, compared to that for people cited only once.
This is true for the numbers of citations, as the median value for
unique-two authors (863) is a bit over twice the median value for
unique-one authors (381). This, however, does not explain the
difference in cite/pap ratio between unique-one and unique-two
astronomers.  Moreover, examination of the astronomer names in the
unique-two subset shows that many are well-known astronomers whose
papers are generally highly regarded.  

We are forced to conclude that real differences exist between the
citation patterns for unique-two astronomers versus those for
unique-one astronomers.  Why this is so is open to speculation.  Is it
possible the confidence a person has in giving two sets of initials
for her/his papers is related to how useful the papers are for
astronomy?  Or is this more a product that a person collaborating with
other individuals puts his/her on the paper without verifying how
her/his name is given?  Explaining this result is beyond the scope of
this paper, and likely would involve sociologists more than astronomers.

\subsection{Honors versus Citations}

As enumerated earlier (Section 3.3), 237 of the 278 astrophysicists in
the honored list are also included in the citation list.  Following
what was done for the chemist list by Dr. Pendlebury, we concentrate
on how the most-cited people in each subset of the ACD (unique-one,
unique-two, confused-named) have been honored; i.e., those astronomers
listed in Table~2.  This list includes 54 astronomers in the
unique-one category (top 1.2\%); 22 astronomers in the unique-two
category (top 6.2\%) and 54 astronomers in the confused-name category
(top 3.6\%), of which 12 can be placed in the Cg category.

The honors received by each person are coded after the citation
information by letter code with the last two digits of the date
awarded by the code (many of which are seen in Table~2, all of which
are used in the full table of honored individuals that will be made
electronically available): International: NOB = Nobel Prize; National,
NAS = member of the National Academy of Sciences; R = Henry Norris
Russell Prize, H = Heinemann Prize.  

We will view the relationship between honors and citations from the
direction of first selecting on citations.  This is because
name-confusion fuzzes the numbers to which we might compare for
unique-two astronomers versus the other two categories.  Until such
name confusion is settled, the only valid statistics go from citations
to honors, not the other way around.  A glance at Table~2 pretty much
tells the whole story.  While only 1 of 22 Unique-Two astronomers
(4.6\%), and only 2 of 42 full-confused-named astronomers (4.9\%) have 
been given lifetime honors, 17 of 54 unique-one astronomers (31.5\%) and
four of 12 Cg astronomers (33.3\%) have been so-honored.  Similarly, 
of 59 non-U.S. astronomers, 3 (5.1\%) have been so-honored, compared
to 21 of 71 U.S. astronomers (29.6\%) so-honored.  Finally, while 
24 of the 130 astronomers/names cited 3000 times or more are so-honored
(18.5\%), only 78 of the 6328 astronomers/names (1.2\%) are so-honored.

The data in Table~2 and in the ACD tell the following story: To first
order, the number of citations your papers receive are, indeed, reasonably
correlated with whether you are honored by your peers.  To second
order, whether or not an individual with high numbers of citations has
been honored to date is a function of several variables: into which
citation subset (unique-one, unique-two or confused) her/his name
falls; how many citations that person has; and from what country that
person mostly does his/her work.  This last effect is understandable,
as most of the honors listed in Table~2 are given by the Amercian
Astronomical Society and the U.S. National Academy of Sciences.

\section{Summary and A Modest Proposal}

An Astronomy Citation Database (ACD) has been assembled,
which gives the number of citations, papers cited and the cite/pap
ratio for 6331+ astronomers for a 16.5 year shapshot of time.  These
data correspond to citation information assembled by the Institute for
Science Information (ISI) for astronomers and physicists for papers
published in the years 1981-1997.0, as cited in papers published in
1981-1997.5.  The data for the astronomy list was assembled from two
databases given to this writer by the ISI.  One list (P\&A-100) contains
citation data for 62,813 physics and astronomy ISI ``names'' cited 100
or more times during the specific intervals.  The other list
(AST-top-papers) contains both citation data for the 200 most-cited
papers published each year in astronomy during 1981-1996, as well as
for the astronomy-related citations for 5,035 astronomer names from
those papers.

The databases given to this author by the ISI give full citation
credit to each author for each of her/his papers, whether first author
or not, whether multi-authored or not.  This is also the methodology
used for ISI's Web of Science.  While the methodology of the ISI in
this regard has not changed over the years, how it presents its data
has.  In particular, users familiar with the hard-copy Science
Citation Index will note that only papers on which you are a first
author are listed in the Author section, while all papers on which you
were an author (first or not) are given the Source section.  As such,
readers should be aware that the ISI Web of Science now gives full
credit for the citations for all papers on which each of us is an
author, but only for refereed journals listed from their list of
sampled journals (cf. Table~1).

In order to make sense of these citation data, this writer had to engage
in a series of laborious, time-consuming tasks.  These tasks also involved
discovering and correcting for a number of biases in the original data
lists, some of which are inherent to any electronically-assembled
database.  

The main file for ACD is divided into three subsets: 4617
``unique-one'' astronomers (those whose names are uniquely identified
with individuals); 357 ``unique-two'' astronomers (those whose last
names are cited with two or more sets of first initials); and 1484
``confused-named'' astronomers (those with names and initials that are
confused with those of other astronomers and/or physicists).  The use
of the word ``unique'' to name two of the subsets refers to the fact
that these are names singly assigned to an individual astronomer.
Such is not the case for the ``confused'' ISI names, which are
associated with two or more individuals in astronomy and/or physics.
Due to name confusion, it is likely than many astronomers are listed
twice in the ACD, either two times in the confused-named list, or once
in the unique-one list and once in the confused-named list.  

Two other files are provided with the ACD.  One is an ``honorable
mention'' list of 173 names for those astronomers whose names are in
the AST-top-papers list but not in the P\&A-100 list, and which have
less than 100 citations from the AST-top-papers list.  The other
is the ``honors'' list, which correlates the citation data information
from the main database with the honors 278 individuals have received for
their astronomical work.

A list of the 10 most-cited papers per year, from 1981 to 1996 is
provided in this paper (but not in the electronic database).
Comparison of the authors' names on the 10 most-cited papers to those
names most cited overall shows a good correspondence.  Moreover, the
papers in the 10 most-cited-per-year list are of a wide range of paper
type (e.g., review, data, theory, observation) and are all known to be
of high quality.  Hence, the old shibboleth that one can get many
citations from publising papers with wrong results is shown to be the
myth that it is.

The errors for citations for individuals in the ACD are both random
and systematic.  Random errors exist in proportion (estimated to be
4\%) to the number of citations, owing to the way in which the ISI
compiles citations.  Systematic errors include errors of omission and
comission.  Errors in citations owing to variance in self-citations
among astronomers can also affect the statistics at the $\sim 6\%$
level, if one chooses to apply such a criterion to these data.  The
likely most-egregious error in this database in the eyes of some of
the readers of this paper is the use of full citations for each author
of a paper.  Tests using the papers of top-cited authors in the
AST-top-papers database shows that no way we can think of to calculate
citations for authors will give the same results.  The plain fact is,
however, whether one agrees with using full citations or not for
astronomers, this is what the ISI gives us in the Web of Science, as
well as in the P\&A-100 and AST-top-papers lists.

We show that, due to the manner in which citations are assembled by
the Institute for Science Information, which maintains the science
citations for our field of study, we hurt the impact of our papers if
we publish many papers in meeting proceedings, or if our name is confused
with those of others.

Two more of our findings are of specific sociological interest.  The
first is that astronomers who divide their papers among two or more
first sets of initials on average publish more papers, have more
citations, and have a significantly higher citations/paper ratio than
astronomers who publish under just one set of initials.  

The second is that there is a very good, but not perfect,
correspondence of a person being near the top of the citation list and
the chance that person has been honored by her/his peers in
U.S. astronomy for life-time achievement.  Closer examination of this
good correlation reveals that assigning honors to our peers is as
human an enterprise as any that we do.  The evidence asuggests that
anything we do, intended or not, to blur the focus of our papers to
our peers has a measurable effect on how we are honored by them.  Such
blurring effects can come in several forms: authoring papers using two
or more sets of initials; having a name confused with those of others;
publishing your papers predominantly in non--U.S.--based journals or
meeting proceedings.  Other means of blurring the focus of our papers
(having high citations, but low cite/pap ratio; mostly publishing in
large groups or with more well-known authors; publishing in several
different scientific fields) also likely exist.

In a paper devoted to analyzing the citation data for astronomers, it
is relevent to note that citation statistics give us but one view of
the impact of individuals on our science.  This is best evidenced by
the most recent Nobel Prize awarded to astronomers, that to JH Taylor
and RA Hulse in 1993.  Where these two individuals stand in the
citation list has little relationship to their scientific impact on
our field.

The two ISI databases used in this paper were generated with ISI
software and given to this author expressly for the analysis done in
this paper.  These databases are different from those we can access
via the Web in three ways: ability to specify specific ranges of years
for cited and citing papers; kinds of papers cited, and number of
citations attributed to each author.  The idiosyncracies discovered
in the course of this survey of the use of current web-based database
for the bibliography of astronomers are detailed.

The full ACD (all three files) will be made available via anonymous
ftp from samuri.la.asu.edu, as well as through the Astrophysics Data
Center.  The top ten papers cited for each year between 1981-1996 are
given in this paper.  The full top-200 cited papers database produced
by the ISI is proprietary (termed ``High-Impact Papers in Astronomy'')
and is available for a fee separately from the ISI.

Since the ACD is a database about the accomplishments of people, any
error in the ACD is a serious error.  As such, readers are encouraged
to contact this writer if errors of omission or comission are found in
the ACD.  Those appropriate modifications to the ACD that should be
made, will be made.  It is hoped that with the help of readers, a full
{\it unconfused}, ACD can be eventually made of cited papers during
1981--1997.5.  Such a list can then act as a baseline against which
future investigations of this kind may be made.  It also hoped that
the lessons learned in this paper about the idiosyncracies of the
various electronic databases will aid others in their own searches.

Towards this end, I end this paper by unashamedly borrowing from
Jonathan Swift in putting forward a ``modest proposal'' for
eliminating name confusion in our field.  We are used to having social
security numbers, university ID numbers, shopper ID numbers; each of
us is now various numbers in various databases.  If this is so, why do
we not assign a ``publishing ID number'' (PID for short) for each
person who publishes a paper in our journals?  I suggest that this be
done, and we start by assigning PIDs to all astronomers who {\it have}
published papers in the journals in the past.  The ACD could be a
start, but only a start, as to solve name confusion the process must
work iteratively among the ISI, the individuals involved and the ACD.
The PID number would then be carried by each journal in the author
list (but not necessarily printed out for each paper).

If we, and other scientists in other fields of study, are interested
in having a true, honest assessment of citation data for astronomers
as a function of time, then the problem of name confusion should be,
and can be, solved.

\vspace*{3mm}

This work could not have been done without the active participation
and cooperation of Dr. David Pendlebury of the ISI, who supplied the
two ISI datasets used in this study, and who also did a very careful
reading of the first draft of this paper.  This author owes
Dr. Pendlebury much thanks.  Conversations with Helmut Abt, Sandra
Faber and Anne Cowley were also helpful in writing this paper.
My thanks to Arnab Choudhuri for pointing me in the direction of the
online Astronomy Society of India directory.  Insightful comments 
from the editor, Bob Milkey, also greatly helped the presentation of 
this paper.

\newpage

\centerline{Figure Caption}

Figure 1: The histograms for the average citations/paper (cite/pap)
for the unique-one (4617 names), unique-two (357 names) and
confused-named (1484 names) subsets in the Astronomy Citation
Database. a) The cite/pap ratio is binned by 5, and the histogram
values are in terms of the fraction of the astronomers in each subset,
falling in each cite/pap bin.  Note the higher median value of
cite/pap, and longer tail towards higher values of cite/pap for the
unique-two astronomers as opposed to the unique-one astronomers. 
b) Numbers of citations are binned by 100.  Histogram values as in
Figure~1a.  Note, as with Figure~1a, the higher median value of
citations, and longer tail towards higher numbers of citations for the
unique-two astronomers as opposed to the unique-one astronomers. c)
Numbers of papers cited for the unique-one, unique-two, and confused
names in our astronomer sample. The number of papers cited is binned
by 10, and the histogram values are in terms of the fraction of the
astronomers in each list, falling in each paper-cited bin.  Note the
higher median value of numbers of papers cited for the unique-two
astronomers relative to those for the unique-one astronomers, and the
longer tails towards high values of papers cited for both unique-two
and confused-named astronomers.


\begin{references}

\reference{} Abt, H.A. 1980, PASP 92, 249

\reference{} Abt, H.A. 1981a, PASP 93, 207

\reference{} Abt, H.A. 1981b, PASP 93, 269

\reference{} Abt, H.A. 1982, PASP 94, 213

\reference{} Abt, H.A. 1983, PASP 95, 113

\reference{} Abt, H.A. 1984a, PASP 96, 563

\reference{} Abt, H.A. 1984b, PASP 96, 746

\reference{} Abt, H.A. 1985, PASP 97, 1050

\reference{} Abt, H.A. 1987a, PASP 99, 439

\reference{} Abt, H.A. 1987b, PASP 99, 1329

\reference{} Abt, H.A. 1988a, PASP 100, 506

\reference{} Abt, H.A. 1988b, PASP 100, 1567

\reference{} Abt, H.A. 1989, PASP 101, 505

\reference{} Abt, H.A. 1990a, PASP 102, 368

\reference{} Abt, H.A. 1990b, Current Contents 33/39, 7

\reference{} Abt, H.A. 1992a, PASP 104, 235

\reference{} Abt, H.A. 1992b, Scientometrics 24, 441

\reference{} Abt, H.A. 1996, PASP 108, 1059

\reference{} Abt, H.A. 1998a, PASP 110, 210

\reference{} Abt, H.A. 1998b, Nature 395, 756

\reference{} Abt, H.A. \& Zhou, H. 1996, PASP 108, 375

\reference{} Davoust, E \& Schmadel, L.D. 1987, PASP 99, 700

\reference{} Davoust, E \& Schmadel, L.D. 1992, Scientometrics 22, 9

\reference{} de Vaucouleurs, G., de Vaucouleurs, A., Corwin, H. C. G., Jr.,
  Buta, R. J., Paturel, G. \& Fouqu\'{e}, P. 1991, Third Reference Catalog of
  Bright Galaxies (New York: Springer-Verlag)

\reference{} Garfield, E. \& Welljams-Dorof, A. 1992, Theoretical
  Medicine, 13, 117

\reference{} Girard, R. \& Davoust, E. 1997, A\&A 323, 1

\reference{} Hoffleit, D. 1982, Bright Star Catalog (New Haven: 
  Yale University)

\reference{} Sandage, A. \& Tammann, G.A. 1981, A Revised Shapley-Ames 
  Catalog of Bright Galaxies, (Washington: Carnegie Institution of Washington)

\reference{} Sandage, A. \& Tammann, G.A. 1987, A Revised Shapley-Ames 
  Catalog of Bright Galaxies, (Washington: Carnegie Institution of Washington)

\reference{} Trimble, V. 1985, QJRAS 26, 40 

\reference{} Trimble, V. 1986a, PASP 98, 1347

\reference{} Trimble, V. 1986b, Czechosl. J. Phys. 36B, 175

\reference{} Trimble, V. 1988, PASP 100, 646

\reference{} Trimble, V. 1991, Scientometrics 20, 71

\reference{} Trimble, V. 1993a, QJRAS 34, 235 

\reference{} Trimble, V. 1993b, QJRAS 34, 301 

\reference{} Trimble, V. 1996, Scientometrics 36, 237

\reference{} White, J.C. II 1992, PASP 104, 472

\end{references}
\end{document}

\setcounter{page}{35}
\footnotesize
\begin{deluxetable}{rrrlrlrr}
\tablenum{3}
\tablecolumns{8}
\tablewidth{0pt}
\tablecaption{Top 10 Cited Papers, By Year, 1981--1996}
\tablehead{
\colhead{No.} & \colhead{Abs} & \colhead{\# in} & \colhead{Authors} & \colhead{Year} & \colhead{Journal}
& \colhead{Vol} & \colhead{Page} \\
\colhead{Cites} &  \colhead{Rank} & \colhead{Yr} &  \colhead{} & \colhead{Publ} & \colhead{} &
\colhead{} & \colhead{} \\
\colhead{1} &  \colhead{2} & \colhead{3} &  \colhead{4} & \colhead{5} & \colhead{6} & \colhead{7} &
\colhead{8} \\
}
\startdata
&&&&&&& \nl
 627 & 13 & 1& Pringle, JE                                                  &1981  & ARA\&A   &   19& 137 \nl
 603 & 16 & 2& Vernazza, JE, Avrett, EH, \& Loeser, R                       &1981  & ApJS     &   45& 635 \nl
 473 & 30 & 3& Renzini, A \& Voli, M                                        &1981  & A\&A     &   94& 175 \nl
 466 & 33 & 4& Kutner, ML \& Ulrich, BL                                     &1981  & ApJ      &  250& 341 \nl
 431 & 42 & 5& Ostriker, JP \& Cowie, LL                                    &1981  & ApJL     &  243& 127 \nl
 398 & 51 & 6& Larson, RB                                                   &1981  & MNRAS    &  194& 809 \nl
 395 & 53 & 7& Baldwin, JA, Phillips, MM, Terlevich, R                      &1981  & PASP     &   93&   5 \nl
 382 & 55 & 8& Kwan, J \& Krolik, JH                                        &1981  & ApJ      &  250& 478 \nl
 372 & 60 & 9& Fisher, JR \& Tully, RB                                      &1981  & ApJS     &   47& 139 \nl
 370 & 62 &10& Vanderhucht, KA, Conti, PS, Lundstrom, I, \& Stenholm, B     &1981  & SSRv     &   28& 227 \nl
&&&&&&& \nl
1100 &  1 & 1& Burstein, D \& Heiles, C                                     &1982  & AJ       &   87&1165 \nl
 634 & 12 & 2& Elias, JE, Frogel, JA, Matthews, K \& Neugebauer, G.         &1982  & AJ       &   87&1029 \nl
 449 & 38 & 3& Scargle, JD                                                  &1982  & ApJ      &  263& 835 \nl
 419 & 46 & 4& Graham, JA                                                   &1982  & PASP     &   94& 244 \nl
 362 & 65 & 5& Sandage, A                                                   &1982  & ApJ      &  252& 553 \nl
 338 & 74 & 6& Malkan, MA \& Sargent, WLW                                   &1982  & ApJ      &  254&  22 \nl
 325 & 79 & 7& Condon, JJ, Condon, MA, Gisler, G \& Puschell, JJ            &1982  & ApJ      &  252& 102 \nl
 314 & 85 & 8& Blandford, RD \& Payne, DG                                   &1982  & MNRAS    &  199& 883 \nl
 311 & 89 & 9& Smith, BA, et al.                                            &1982  & Sci      &  215& 504 \nl
 297 &106 &10& Piccinotti, G, et al.                                        &1982  & ApJ      &  253& 485 \nl
&&&&&&& \nl
 826 &  4 & 1& Landolt, AU                                                  &1983  & AJ       &   88& 853 \nl
 726 &  7 & 2& Iben, I, Jr \& Renzini, A                                    &1983  & ARA\&A   &   21& 271 \nl
 696 &  8 & 3& Davis, M \& Peebles, PJE                                     &1983  & ApJ      &  267& 465 \nl
 682 &  9 & 4& Morrison, R \& McCammon, D                                   &1983  & ApJ      &  270& 119 \nl
 474 & 28 & 5& Bally, J \& Lada, CJ                                         &1983  & ApJ      &  265& 824 \nl
 468 & 32 & 6& Schmidt, M \& Green RF                                       &1983  & ApJ      &  269& 352 \nl
 463 & 34 & 7& Bahcall, NA \& Soniera, RM                                   &1983  & ApJ      &  270&  20 \nl
 457 & 37 & 8& Kennicutt, RC, Jr                                            &1983  & ApJ      &  272&  54 \nl
 445 & 39 & 9& Huchra, JP, Davis, M, Latham, D, \& Tonry, J                 &1983  & ApJS     &   52&  89 \nl
 440 & 40 &10& Hildebrand, RH                                               &1983  & QJRAS    &   24& 267 \nl
&&&&&&& \nl
 617 & 14 & 1& Burstein, D \& Heiles, C                                     &1984  & ApJS     &   54&  33 \nl
 576 & 17 & 2& Yang, J, Turner, MS, Schramm, DN, Steigman, G, \& Olive, KA  &1984  & ApJ      &  281& 493 \nl
 527 & 21 & 3& Patterson, J                                                 &1984  & ApJS     &   54& 443 \nl
 516 & 22 & 4& Leger, A \& Puget, JL                                        &1984  & A\&A     &  137&   5 \nl
 462 & 35 & 5& Zinn, R \& West, MJ                                          &1984  & ApJS     &   55&  45 \nl
 461 & 36 & 6& Blumenthal, GR, Faber, SM, Primack, JR, Rees, MJ             &1984  & Natur    &  311& 517 \nl
 434 & 41 & 7& Jones, C \& Forman, W                                        &1984  & ApJ      &  276&  38 \nl
 408 & 48 & 8& Bridle, AH, Perley, RA                                       &1984  & ARA\&A   &   22& 319 \nl
 405 & 49 & 9& Rees, MJ                                                     &1984  & ARA\&A   &   22& 471 \nl
 359 & 68 &10& Kaiser, N                                                    &1984  & ApJL     &  284&   9 \nl
&&&&&&& \nl
 980 &  2 & 1& Draine, BT \& Lee, HM                                        &1985  & ApJ      &  285&  89 \nl
 642 & 10 & 2& Davis, M, Efstathiou, G, Frenk, CS, \& White, SDM            &1985  & ApJ      &  292& 371 \nl
 547 & 18 & 3& Rieke, GH \& Lebofsky, MJ                                    &1985  & ApJ      &  288& 618 \nl
 532 & 20 & 4& Antonucci, RRJ \& Miller, JS                                 &1985  & ApJ      &  297& 621 \nl
 495 & 25 & 5& Zinn, R                                                      &1985  & ApJ      &  292& 424 \nl
 474 & 28 & 6& Lada, CJ                                                     &1985  & ARA\&A   &   23& 267 \nl
 412 & 47 & 7& VandenBerg, DA                                               &1985  & ApJS     &   58& 711 \nl
 404 & 50 & 8& Boesgaard, AM \& Steigman, G                                 &1985  & ARA\&A   &   23& 319 \nl
 377 & 56 & 9& VandenBerg, DA \& Bell, RA                                   &1985  & ApJS     &   58& 561 \nl
 371 & 61 &10& Knapp, GR \& Morris, M                                       &1985  & ApJ      &  292& 640 \nl
&&&&&&& \nl
 769 &  6 & 1& Bardeen, JM, Bond, JR, Kaiser, N, \& Szalay, AS              &1986  & ApJ      &  304&  15 \nl
 500 & 24 & 2& deLapparent, V, Geller, MJ, Huchra, JP                       &1986  & ApJL     &  302&   1 \nl
 481 & 27 & 3& Mikheyev, SP \& Smirnov, AY                                  &1986  & NUOV CIM C &  9&  17 \nl
 376 & 57 & 4& Woosley, SE \& Weaver, TA                                    &1986  & ARA\&A   &   24& 205 \nl
 356 & 70 & 5& Barnes, J \& Hut, P                                          &1986  & Natur    &  324& 446 \nl
 318 & 82 & 6& Dekel, A \& Silk, J                                          &1986  & ApJ      &  303&  39 \nl
 291 &113 & 7& Green, RF, Schmidt, M, \& Liebert, J                         &1986  & ApJS     &   61& 305 \nl
 288 &115 & 8& Paczynski, B                                                 &1986  & ApJ      &  304&   1 \nl
 282 &120 & 9& Aaronson, M, et al.                                          &1986  & ApJ      &  302& 536 \nl
 265 &145 &10& Horne, K                                                     &1986  & PASP     &   98& 609 \nl
&&&&&&& \nl
 636 & 11 & 1& Hedin, AE                                                    &1987  & J GEO R-S P &92&4649 \nl
 606 & 15 & 2& Shu, FM, Adams, FC, Lizano, S                                &1987  & ARA\&A    &  25&  23 \nl
 374 & 58 & 3& Adams, FC, Lada, CJ, Shu, FH                                 &1987  & ApJ       & 312& 788 \nl
 367 & 63 & 4& Heppner, JP \& Maynard, NC                                   &1987  & J GEO R-S P &92&4467 \nl
 309 & 92 & 5& Dame, TM et al.                                              &1987  & ApJ       & 322& 706 \nl
 306 & 94 & 6& Veilleux, S \& Osterbrock, DE                                &1987  & ApJS      &  63& 295 \nl
 283 &119 & 7& Hewitt, A \& Burbidge, G                                     &1987  & ApJS      &  63&   1 \nl
 280 &124 & 8& Tsyganenko, NA                                               &1987  & P\&SS     &  35&1347 \nl
 275 &130 & 9& Blake, GA, Sutton, EC, Masson, CR \& Phillips, TG            &1987  & ApJ       & 315& 621 \nl
 274 &132 &10& Maeder, A \& Meynet, G                                       &1987  & A\&A      & 182& 243 \nl
&&&&&&& \nl
 825 &  5 & 1& Stetson, PB                                                  &1988  & PASP      &  99& 191 \nl
 428 & 43 & 2& Tyson, JA                                                    &1988  & AJ        &  96&   1 \nl
 421 & 44 & 3& Lyndenbell, D, et al.                                        &1988  & ApJ       & 326&  19 \nl
 420 & 45 & 4& Sanders, DB, et al.                                          &1988  & ApJ       & 325&  74 \nl
 318 & 82 & 5& Maeder, A \& Meynet, G                                       &1988  & A\&AS     &  76& 411 \nl
 294 &109 & 6& Efstathiou, G, Ellis, RS, \& Peterson, BA                    &1988  & MNRAS     & 232& 431 \nl
 271 &137 & 7& Broadhurst, TJ, Ellis, RS, \& Shanks, T                      &1988  & MNRAS     & 235& 827 \nl
 267 &142 & 8& Bertout, C, Basri, G, \& Bouvier, J                          &1988  & ApJ       & 330& 350 \nl
 260 &154 & 9& van Dishoeck, EF, Ewine, F, \& Black, JH                     &1988  & ApJ       & 334& 771 \nl
 230 &205 &10& Lightman, AP \& White, TR                                    &1988  & ApJ       & 335&  57 \nl
&&&&&&& \nl
 485 & 26 & 1& Cardelli, JA, Clayton, GC, \& Mathis, JS                     &1989  & ApJ       & 345& 245 \nl
 361 & 66 & 2& Barthel, PD                                                  &1989  & ApJ       & 336& 606 \nl
 352 & 72 & 3& Wheeler, JC, Sneden, C, \& Truran, JW, Jr                    &1989  & ARA\&A    &  27& 279 \nl
 337 & 76 & 4& Abell, GO, Corwin, HG, Jr, \& Olowin, RP                     &1989  & ApJS      &  70&   1 \nl
 314 & 85 & 5& Geller, MJ \& Huchra, JP                                     &1989  & Sci       & 246& 897 \nl
 305 & 95 & 6& Turner, TJ \& Pounds, KA                                     &1989  & MNRAS     & 240& 833 \nl
 299 &102 & 7& Kennicutt, RC, Jr                                            &1989  & ApJ       & 344& 685 \nl
 298 &103 & 8& Turner, MJL, et al.                                          &1989  & PASJ      &  41& 345 \nl
 264 &149 & 9& Taylor, JH \& Weisberg, JM                                   &1989  & ApJ       & 345& 434 \nl
 263 &151 &10& Tsyganenko, NA                                               &1989  & P\&SS     &  37&   5 \nl
&&&&&&& \nl
 373 & 59 & 1& Beckwith, SVW, Sargent, AI, Chini, RS, \& Guesten, R         &1990  & AJ        &  99& 924 \nl
 322 & 80 & 2& Broadhurst, TJ, Ellis, RS, Koo, DC, \& Szalay, AS            &1990  & Natur     & 343& 726 \nl
 303 & 99 & 3& Maddox, SJ, Efstathiou, G, Sutherland, WJ, \& Loveday, J     &1990  & MNRAS     & 242&  43 \nl
 293 &110 & 4& Mathis, JS                                                   &1990  & ARA\&A    &  28&  37 \nl
 273 &135 & 5& Mather, JC, et al.                                           &1990  & ApJL      & 354&  37 \nl
 265 &145 & 6& Dickey, JM \& Lockman, FJ                                    &1990  & ARA\&A    &  28& 215 \nl
 239 &188 & 7& Pounds, KA, Nandra, K, Stewart, GC, George, GC \& Fabian, AC &1990  & Natur     & 344& 132 \nl
 238 &189 & 8& Gioia, IM, et al.                                            &1990  & ApJS      &  72& 567 \nl
 217 &243 & 9& Colless, M, Ellis, RS, Taylor, K, \& Hook, RN                &1990  & MNRAS     & 244& 408 \nl
 215 &246 &10& Maeder, A                                                    &1990  & A\&AS     &  84& 139 \nl
&&&&&&& \nl
 541 & 19 & 1& Walker, TP, Steigman, G, Kang, HS, Schramm, DM, \& Olive, KA &1991  & ApJ       & 376&  51 \nl
 281 &122 & 2& Morton, DC                                                   &1991  & ApJS      &  77& 119 \nl
 249 &175 & 3& Saunders, W, Frenk, C, Rowan-Robinson, M, Lawrence, A, Efstathiou, G&1991 &Natur& 349&  32 \nl
 235 &196 & 4& Stocke, JT, et al.                                           &1991  & ApJS      &  76& 813 \nl
 227 &213 & 5& Duquennoy, A \& Mayor, M                                     &1991  & A\&A      & 248& 485 \nl
 227 &213 & 5& Tsuneta, S, et al.                                           &1991  & SoPh      & 136&  37 \nl
 212 &257 & 7& Lilly, SJ, Cowie, LL, \& Gardner, JP                         &1991  & ApJ       & 369&  79 \tablebreak
 211 &259 & 8& Balbus, SA \& Hawley, JF                                     &1991  & ApJ       & 376& 214 \nl
 179 &378 & 9& Clavel, J, et al.                                            &1991  & ApJ       & 366&  64 \nl
 173 &413 &10& George, IM \& Fabian, AC                                     &1991  & MNRAS     & 249& 352 \nl
&&&&&&& \nl
 903 &  3 & 1& Smoot, GF, et al.                                            &1992  & ApJL      & 396&   1 \nl
 510 & 23 & 2& Landolt, AU                                                  &1992  & AJ        & 104& 340 \nl
 473 & 30 & 3& Schaller, G, Schaerer, D, Meynet, G, \& Maeder, A            &1992  & A\&AS     &  96& 269 \nl
 398 & 51 & 4& Stark, AA, et al.                                            &1992  & ApJS      &  79&  77 \nl
 309 & 92 & 5& Rogers, FJ \& Iglesias, CA                                   &1992  & ApJS      &  79& 507 \nl
 292 &111 & 6& Wright, EL, et al.                                           &1992  & ApJL      & 396&  13 \nl
 277 &127 & 7& Meegan, CA, et al.                                           &1992  & Natur     & 355& 143 \nl
 220 &234 & 8& Iglesias, CA, Rogers, FJ, \& Wilson, BG                      &1992  & ApJ       & 397& 717 \nl
 219 &236 & 9& Loveday, J, Peterson, BA, Efstathiou, G, \& Maddox, SJ       &1992  & ApJ       & 390& 338 \nl
 214 &250 &10& Jacoby, GH, et al.                                           &1992  & PASP      & 104& 599 \nl
&&&&&&& \nl
 297 &106 & 1& Antonucci, R                                                 &1993  & ARA\&A    &  31& 473 \nl
 278 &126 & 2& Alcock, C, et al.                                            &1993  & Natur     & 365& 621 \nl
 275 &130 & 3& Edvardsson, B, et al.                                        &1993  & A\&A      & 275& 101 \nl
 261 &153 & 4& Taylor, JH, Manchester, N, \& Lyne, AG                       &1993  & ApJS      &  88& 529 \nl
 215 &246 & 5& Bruzual, G \& Charlot, S                                     &1993  & ApJ       & 405& 538 \nl
 190 &319 & 6& White, SDM, Navarro, JF, Evrard, AE, \& Frenk, CA            &1993  & Natur     & 366& 429 \nl
 186 &338 & 7& Taylor, JH \& Cordes, JM                                     &1993  & ApJ       & 411& 674 \nl
 172 &416 & 8& Pounds, KA, et al.                                           &1993  & MNRAS     & 260&  77 \nl
 164 &469 & 9& Klypin, A, Holtzman, J, Primack, J, \& Regos, E              &1993  & ApJ       & 416&   1 \nl
 146 &612 &10& White, SDM,  Efstathiou, G, \& Frenk, CA                     &1993  & MNRAS     & 262&1023 \nl
&&&&&&& \nl
 242 &184 & 1& Freedman, WL, et al.                                         &1994  & Natur     & 371& 757 \nl
 241 &185 & 2& Tanaka, Y, Inoue, H, \& Holt, SS                             &1994  & PASL      &  46&  37 \nl
 186 &338 & 3& Lyne, AG \& Lorimer, DR                                      &1994  & Natur     & 369& 127 \nl
 166 &454 & 4& Mather, JC, et al.                                           &1994  & ApJ       & 420& 439 \nl
 158 &512 & 5& Fichtel, CE, et al.                                          &1994  & ApJS      &  94& 551 \nl
 153 &545 & 6& D'Antona, F \& Mazzitelli, I                                 &1994  & ApJS      &  90& 467 \nl
 153 &545 & 6& Songaila, A, Cowie, LL, Hogan, CJ,\& Rugers, M               &1994  & Natur     & 368& 599 \nl
 152 &560 & 8& Peacock, JA \& Dodds, SJ                                     &1994  & MNRAS     & 267&1020 \nl
 141 &660 & 9& Nandra, K \& Pounds, KA                                      &1994  & MNRAS     & 268& 405 \nl
 139 &693 &10& Bennett, CL, et al.                                          &1994  & ApJ       & 436& 423 \nl
&&&&&&& \nl
 134 &745 & 1& Von Montigny, C, et al.                                      &1995  & ApJ       & 440& 525 \nl
 128 &822 & 2& Copi, CJ, Schramm, DN, \& Turner, MS                         &1995  & Sci       & 267& 192 \nl
 127 &837 & 3& Hotlzman, JA, et al.                                         &1995  & PASP      & 107&1065 \nl
 127 &837 & 3& Thompson, DJ, et al.                                         &1995  & ApJS      & 101& 259 \nl
 110&1189 & 5& Miyoshi, M, et al.                                           &1995  & Natur     & 373& 127 \nl
 108&1239 & 6& Leitherer, C \& Heckman, TM                                  &1995  & ApJS      &  96&   9 \nl
  98&1484 & 7& Mayor, M \& Queloz, D                                        &1995  & Natur     & 378& 355 \nl
  93&1638 & 8& Lilly, SJ, Tresse, L, Hammer, F, Crampton, D, \& Le Fevre, O &1995  & ApJ       & 455& 108 \nl
  92&1668 & 9& Woosley, SE \& Weaver, TA                                    &1995  & ApJS      & 101& 181 \nl
  87&$\ldots$ &10& Timmes, FX, Woosley, SE, \& Weaver, TA                   &1995  & ApJS      &  98& 617 \nl
&&&&&&& \nl
  80&$\ldots$ & 1& Steidel, CC, Giavalisco, M, Pettini, M, Dickinson, M, Adelberger, KL&1996 &ApJL    &   462 & 17  \nl
  66&$\ldots$ & 2& Tytler, D, Fan, X-M, \& Burles, S                                   &1996 &Natur   &   381 &207  \nl
  47&$\ldots$ & 3& Ellis, RS, Colless, M, Broadhurst, T, Heyl, J \& Glazebrook, K      &1996 &MNRAS   &   280 &235  \nl
  47&$\ldots$ & 3& Rugers, M \& Hogan, CJ                                              &1996 &ApJL    &   459 &  1  \nl
  45&$\ldots$ & 5& Narayan, R, McClintock, JE, \& Yi, I                                &1996 &ApJ     &   457 &821  \nl
  44&$\ldots$ & 6& Babul, A \& Ferguson, HC                                            &1996 &ApJ     &   458 &100  \nl
  43&$\ldots$ & 7& Marcy, GW \& Butler, RP                                             &1996 &ApJ     &   464 &147  \nl
  42&$\ldots$ & 8& Navarro, JF, Frenk, CS, White, SDM                                  &1996 &ApJ     &   462 &563  \nl
  42&$\ldots$ & 8& Williams, RE, et al.                                                &1996 &AJ      &   112&1335  \nl
  40&$\ldots$ &10& Alcock, C, et al.                                                   &1996 &ApJ     &   461 & 84  \nl
  40&$\ldots$ &10& Chaboyer, B, DeMarque, P, \& Sarajedini, A                          &1996 &ApJ     &   459 &558  \nl
  40&$\ldots$ &10& Hernquist, L, Katz, N, Weinberg, DH, \& Miralda-Escude$'$, J        &1996 &ApJL    &   457 & 51  

\enddata
\end{deluxetable}

\setcounter{page}{32}
\footnotesize
\begin{deluxetable}{llrrrclll}
\tablenum{2}
\tablecolumns{9}
\tablewidth{0pt}
\tablecaption{Citation and Honor Data for Astronomers Cited 3000+ Times}
\tablehead{
\colhead{Last} & \colhead{First} & \colhead{Cite/} & \colhead{No.} & \colhead{No.}
& \colhead{Typ} & \colhead{Hon} & \colhead{Cty} & \colhead{Comments} \\
\colhead{Name} &  \colhead{Init} & \colhead{Pap} &  \colhead{Pap} & \colhead{Cites} & 
\colhead{} & \colhead{} & \colhead{} & \colhead{} \\
\colhead{1} &  \colhead{2} & \colhead{3} &  \colhead{4} & \colhead{5} & 
\colhead{6} & \colhead{7} & \colhead{8} & \colhead{9}
}
\startdata
\multicolumn{9}{c}{Unique-Two Astronomers} \nl
&&&&&&&& \nl
HUCHRA          &     JP  &  51.17 & 160 & 8187 & 2 &NAS93             &     &  J~~~\,+2531~~~30~~~A16~~+223~~~3  \nl
EFSTATHIOU      &     G   &  77.21 &  95 & 7335 & 2 &                  &  UK &  GP~~\,+151~~~~3             \nl
OLIVE           &     KA  &  32.39 & 162 & 5247 & 2 &                  &     &  K~~~~\,+530~~~~6            \nl
SCOVILLE        &     NZ  &  44.87 & 113 & 5070 & 2 &                  &     &  N~~~+1099~~~25               \nl
LINDE           &     AD  &  57.55 &  86 & 4949 & 2 &                  &  Ru &  A~~~+1136~~~41              \nl
MOULD           &     JR  &  28.82 & 164 & 4726 & 2 &                  &  Au &  J~~~\,+2084~~~60~~~A16~~~+28~~~1  \nl
MUSHOTZKY       &     RF  &  31.46 & 138 & 4342 & 2 &                  &     &  R~~~~\,+539~~~26            \nl
LIEBERT         &     J   &  26.48 & 159 & 4211 & 2 &                  &     &  JW~~+253~~~~8               \nl
ELLIS           &     RS  &  35.34 & 119 & 4206 & 2 &                  &  UK &  R~~~~\,+653~~~20~~~A16~~~+28~~~1 \nl
HECKMAN         &     TM  &  37.44 & 110 & 4118 & 2 &                  &     &  T~~~+1168~~~25              \nl
SCHRAMM         &     DN  &  30.00 & 137 & 4110 & 2 &                  &     &  D~~~~\,+100~~~~5            \nl
FRENK           &     CS  &  59.96 &  68 & 4077 & 2 &                  &  UK &  C~~~~\,+360~~~~7            \nl
SMOOT           &     GF  &  37.26 & 105 & 3912 & 2 &                  &     &  G~~~~\,+310~~~15~~~A16~~+643~~~4 \nl
KENNICUTT       &     RC  &  37.32 & 104 & 3881 & 2 &                  &     &  R~~~~\,+147~~~~8~~~A16~~~+32~~~1 \nl
BOTHUN          &     GD  &  28.19 & 127 & 3580 & 2 &                  &     &  G~~~~\,+561~~~10            \nl
WEGNER          &     G   &  23.33 & 143 & 3336 & 2 &                  &     &  GA~~+138~~~~2               \nl
TERLEVICH       &     R   &  35.89 &  91 & 3266 & 2 &                  &  UK &  RJ~\,+1766~~~21~~~A16~~~+28~~~1 \nl
BOGGESS         &     NW  & 127.48 &  25 & 3187 & 2 &                  &     &  N~~~+1027~~~~9~~~A16~~+128~~~1  \nl
STOCKE          &     JT  &  32.76 &  96 & 3145 & 2 &                  &     &  J~~~\,+1018~~~25            \nl
MILEY           &     GK  &  33.30 &  94 & 3130 & 2 &                  &     &  G~~~A16~~+274~~~1 \nl
LUBIN           &     PM  &  75.12 &  41 & 3080 & 2 &                  &     &  P~~~~\,+802~~~17~~~A16~+1035~~~2 \nl
MAYERHASSELWANDER &   HA  &  38.10 &  79 & 3010 & 2 &                  &  Ge &  HA~~A16~~+235~~~3~~~H~~~A16~~+160~~~1 \nl
&&&&&&&& \nl
\multicolumn{9}{c}{Unique-One Astronomers} \nl
&&&&&&&& \nl
FABIAN          &     AC  &  25.38 & 258 & 6549 & 1 &                  &   UK&                              \nl    
SANDAGE         &     A   &  59.36 & 106 & 6292 & 1 &R72, NAS(decl)    &     & +ca:~+1250~~~1~~for RSA      \nl 
SOIFER          &     BT  &  44.67 & 135 & 6030 & 1 &                  &     & ~~~A16~~+274~~~1             \nl        
BAHCALL         &     JN  &  36.34 & 163 & 5923 & 1 &NAS76, H94, R99   &     &                              \nl        
TURNER          &     MS  &  32.44 & 171 & 5547 & 1 &NAS97             &     &                              \nl        
STEINHARDT      &     PJ  &  72.84 &  76 & 5536 & 1 &NAS98             &     &                              \nl        
RUSSELL         &     CT  &  20.50 & 264 & 5413 & 1 &                  &     &  Christopher T               \nl        
BURSTEIN        &     D   &  76.03 &  71 & 5398 & 1 &                  &     &  David                       \nl     
VILENKIN        &     A   &  50.32 &  96 & 4831 & 1 &                  &   Ru&  Alexander                   \nl        
WHITE           &     SDM &  51.06 &  94 & 4800 & 1 &                  &   UK&  Simon, (White SM confused)  \nl        
REES            &     MJ  &  36.50 & 131 & 4782 & 1 &NAS82, H84        &   UK&                              \nl        
LAMBERT         &     DL  &  25.41 & 182 & 4625 & 1 &                  &   UK&                              \nl        
OSTRIKER        &     JP  &  42.19 & 108 & 4557 & 1 &NAS74, R80        &     &                              \nl        
BROWN           &     GE  &  27.40 & 165 & 4521 & 1 &                  &     &  Gerald E Jr                 \nl
FABER           &     SM  &  56.36 &  80 & 4509 & 1 &H85, NAS85        &     &                              \nl        
ROWANROBINSON   &     M   &  41.82 & 105 & 4391 & 1 &                  &   UK&  Michael~~~~A16~~+274~~~1    \nl
GUNN            &     JE  &  48.74 &  88 & 4289 & 1 &NAS77, H88        &     &                              \nl        
MAEDER          &     A   &  60.91 &  70 & 4264 & 1 &                  &   Sw&  Andre                       \nl        
SANDERS         &     DB  &  48.43 &  82 & 3971 & 1 &                  &     &  David B                     \nl
HAWKING         &     SW  &  72.16 &  55 & 3969 & 1 &NAS92             &   UK&                              \nl
SILK            &     J   &  19.90 & 199 & 3961 & 1 &                  &   UK&  Joseph I                    \nl
GELLER          &     MJ  &  41.98 &  92 & 3862 & 1 &NAS92             &     &                              \nl
IBEN            &     I   &  47.47 &  79 & 3750 & 1 &R89, NAS85        &     &  Icko, Jr                    \nl
MATTHEWS        &     K   &  30.22 & 124 & 3747 & 1 &                  &     &  Keith                       \nl
BAME            &     SJ  &  29.75 & 125 & 3719 & 1 &                  &     &  Samuel J                    \nl
DRESSLER        &     A   &  54.63 &  68 & 3715 & 1 &NAS96             &     &  Alan                        \nl
SARGENT         &     WLW &  38.97 &  94 & 3663 & 1 &H91               &     &                              \nl
HEILES          &     C   &  58.23 &  61 & 3552 & 1 &H89, NAS90        &     &  Carl                        \nl
LADA            &     CJ  &  51.42 &  69 & 3548 & 1 &                  &     &                              \nl
HAUSER          &     MG  &  81.79 &  43 & 3517 & 1 &                  &     &  Michael G~~~A16~~+166~~~1   \nl
THADDEUS        &     P   &  38.60 &  91 & 3513 & 1 &NAS87             &     &  Patrick                     \nl
LINSKY          &     JL  &  29.67 & 118 & 3501 & 1 &                  &     &  ~~~A16~~~+32~~~1            \nl
TIELENS         &     AGGM&  34.42 & 101 & 3476 & 1 &                  &  Ne &                              \nl
WOOSLEY         &     SE  &  46.01 &  75 & 3451 & 1 &                  &     &                              \nl
TAMMANN         &     GA  &  79.81 &  43 & 3432 & 1 &                  &  Sw &  +ca:~+1250~~~1~~for RSA     \nl
ROSNER          &     R   &  28.26 & 121 & 3419 & 1 &                  &     &  Robert                      \nl
STEIGMAN        &     G   &  50.55 &  67 & 3387 & 1 &                  &     &  Gary                        \nl
GENZEL          &     R   &  31.65 & 107 & 3387 & 1 &                  &  Ge &  Reinhard                    \nl
SNELL           &     RL  &  35.53 &  95 & 3375 & 1 &                  &     &                              \nl
GOSLING         &     JT  &  24.80 & 136 & 3373 & 1 &                  &     &  John T                      \nl
GURNETT         &     DA  &  22.73 & 145 & 3296 & 1 &                  &     &  Donald A                    \nl
WRIGHT          &     EL  &  50.70 &  64 & 3245 & 1 &                  &     &  ~~~A16~~+244~~~3            \nl
KIRSHNER        &     RP  &  33.24 &  97 & 3224 & 1 &NAS98             &     &  +sp:~~+191~~~33~~Kirschner~~~A16~~~+63~~~1 \nl
BOKSENBERG      &     A   &  30.40 & 105 & 3192 & 1 &                  &  UK &  Alex                        \nl
KANBACH         &     G   &  31.65 & 100 & 3165 & 1 &                  &  Ge &  Gottfried~~~A16~~+127~~~1   \nl
WHITE           &     NE  &  26.15 & 121 & 3164 & 1 &                  &     &  Nicholas E                  \nl
DRAINE          &     BT  &  68.76 &  46 & 3163 & 1 &                  &     &                              \nl
SAVAGE          &     BD  &  31.05 & 101 & 3136 & 1 &                  &     &                              \nl
MAKISHIMA       &     K   &  21.18 & 147 & 3128 & 1 &                  &  Ja &  Kasuo~~~A16~~~+35~~~1       \nl
BALLY           &     J   &  38.10 &  82 & 3124 & 1 &                  &     &  John                        \nl
WEYMANN         &     RJ  &  41.64 &  74 & 3081 & 1 &NAS84             &     &  ~~~~~~~~A16~~+178~~~2       \nl
GEBALLE         &     TR  &  18.98 & 161 & 3056 & 1 &                  &     &                              \nl
BOND            &     JR  &  76.30 &  40 & 3052 & 1 &                  &  Ca &  J Richard                   \nl
VANDENBERG      &     DA  &  63.46 &  48 & 3046 & 1 &                  &  Ca &                              \nl
&&&&&&&& \nl
\multicolumn{9}{c}{Confused--g Astronomers} \nl
&&&&&&&& \nl
DAVIS           &     M   &  70.52 &  95 & 6699 & Cg &NAS91             &     &   mostly Marc, also Michael M, + others \nl
NEUGEBAUER      &     G   &  40.32 & 166 & 6693 & Cg &NAS73, R96        &     &   Gerry, Greg (p), GT (p)     \nl             
KAISER          &     N   &  32.19 & 136 & 4378 & Cg &                  &   UK&   Nick + N (p)                \nl            
SCHMIDT         &     M   &  21.02 & 174 & 3657 & Cg &R78, NAS78        &     &   Maarten, + others           \nl   
INOUE           &     H   &  14.66 & 248 & 3635 & Cg &                  &   Ja&   Hajime, Hiroyuik, + others  \nl
KOYAMA          &     K   &  12.43 & 287 & 3568 & Cg &                  &   Ja&   Katsui, Kazuya, Kouichi, + others \nl
HARTMANN        &     L   &  38.43 &  89 & 3420 & Cg &                  &     &   Lee W, Lawrence (p)        \nl    
TAYLOR          &     JH  &  46.33 &  73 & 3382 & Cg &Nob93, H80, NAS81 &     &   Joseph H, + others~~A16~~+185~~3~Joseph~H \nl    
BENNETT         &     CL  &  40.30 &  83 & 3345 & Cg &                  &     &   Charles L, Charles L (p)   \nl    
GREEN           &     RF  &  34.30 &  92 & 3156 & Cg &                  &     &   Richard F, RF (p)~~A16~~~+44~~1~Richard~F  \nl    
NOMOTO          &     K   &  27.15 & 116 & 3149 & Cg &                  &   Ja&   Ken-Ichi, Kuninori (p)     \nl               
BECHTOLD        &     J   &  44.82 &  68 & 3048 & Cg &                  &     &   Jill, Jeffery J (p)        \nl    
&&&&&&&& \nl
\multicolumn{9}{c}{Confused Astronomers} \nl
&&&&&&&& \nl
SUZUKI          &     T   &   7.58 &1401 &10617 & C &                  &   Ja&   Takao, Tomoharu, Toshihiro, + others \nl
ANDERSON        &     PW  &  80.30 & 130 &10439 & C &                  &     &   PW, + others                \nl
MURPHY          &     DW  & 111.04 &  76 & 8439 & C &                  &     &   David W, DW (p)             \nl            
TANAKA          &     K   &   7.96 &1020 & 8122 & C &                  &   Ja&   4 names, + others~~~~A16~~~+77~~~1 \nl
WISE            &     MB  &  57.44 & 138 & 7927 & C &                  &     &   Mark B, Mark B (phys)       \nl
TAKAHASHI       &     K   &  12.05 & 637 & 7675 & C &                  &   Ja&   many names, + others        \nl
TANAKA          &     Y   &  13.97 & 535 & 7474 & C &NAS98             &   Ja&   Yasuo,Yoshio,Yutaka,+others~~A16~~~+39~1~Yasuo \nl
SATO            &     M   &  10.88 & 607 & 6606 & C &                  &   Ja&   Makoto, Massae, + 12 others \nl            
MURAKAMI        &     T   &  22.46 & 291 & 6537 & C &                  &   Ja&   Tadayoshi, Toshio, + others~~A16~~+210~~~1 \nl
SUZUKI          &     K   &   7.29 & 791 & 5769 & C &                  &   Ja&   many names, + others        \nl
TAKAHASHI       &     T   &   9.00 & 628 & 5650 & C &                  &   Ja&  Takamasa,Toru,Toshikazu,+other~A16~+66~1~Tadayuki \nl
SATO            &     K   &   7.69 & 655 & 5036 & C &                  &   Ja&   Ken-Ichi, + others          \nl
SAITO           &     S   &  13.42 & 365 & 4900 & C &                  &   Ja&   Sumisaburo, Shuji, + others \nl            
GUPTA           &     A   &  16.88 & 275 & 4642 & C &                  &   In&   Anjana, Ajay, Alakananda, Alka, Anshu, + others \nl
SCHWARZ         &     JH  &  79.78 &  55 & 4388 & C &                  &     &   Joseph H, John H (p)        \nl            
TANAKA          &     H   &   7.87 & 548 & 4314 & C &                  &   Ja&   Hidekazu, Haruo, + others   \nl            
SATO            &     N   &  14.12 & 301 & 4251 & C &                  &   Ja&   Naohiro,Naohisa,Nobuaki,+others~~A16~~+60~2  \nl
SHIMIZU         &     T   &  10.12 & 409 & 4140 & C &                  &   Ja&   4 names, + others           \nl
WANG            &     Y   &   8.94 & 456 & 4077 & C &                  &   Ch&   Yi-Ming, Yun, + others      \nl           
ALLEN           &     JW  &  31.29 & 130 & 4068 & C &                  &     &   John W, James W (p)         \nl            
SEN             &     A   &  19.37 & 202 & 3913 & C &                  &   In&   AK, + others \nl
STONE           &     AD  &  80.48 &  48 & 3863 & C &                  &     &   Abraham D, Alfred D (p)     \nl   
SAKURAI         &     T   &  10.43 & 370 & 3858 & C &                  &   Ja&   many names, + others        \nl
KOCH            &     RH  &  29.97 & 127 & 3806 & C &                  &     &   Robert H, Roger H (p)       \nl   
WATANABE        &     T   &   6.02 & 618 & 3722 & C &                  &   Ja&   4 names, + others           \nl
SUZUKI          &     S   &   8.13 & 452 & 3675 & C &                  &   Ja&   many names, + others        \nl
INOUE           &     M   &   7.37 & 490 & 3610 & C &                  &   Ja&   many names, + others        \nl
STEWART         &     GR  &  21.12 & 169 & 3569 & C &                  &     &   Glen R, Gregory R (p)       \nl   
KOYAMA          &     K   &  12.43 & 287 & 3568 & C &                  &   Ja&   Katsui, Kazuya, Kuichi, + others  \nl
MATSUOKA        &     M   &  14.37 & 240 & 3449 & C &                  &   Ja&   Masaru, + others~~~A16~~+132~~~1 \nl
SUZUKI          &     H   &   6.83 & 504 & 3442 & C &                  &   Ja&   many names, + others        \nl
CHEN            &     J   &   8.29 & 415 & 3442 & C &                  &   Ch&   Jian-sheng, James, Jun, + other \nl 
HARTMANN        &     L   &  38.43 &  89 & 3420 & C &                  &     &   Lee W, Lawrence(p)         \nl
COHEN           &     M   &  19.44 & 174 & 3383 & C &NAS84             &     &   Marshall H, Martin, + others \nl
KATO            &     M   &   8.36 & 393 & 3285 & C &                  &   Ja&   Mariko, + others           \nl    
SINGH           &     J   &   8.16 & 402 & 3280 & C &                  &   In&   Jagdev, Jagjit, Jyoti, Jagadish, + others \nl
ODA             &     M   &  25.90 & 126 & 3263 & C &                  &   Ja&   Minoru, Mitsushige, + others \nl
PINES           &     D   &  39.80 &  81 & 3224 & C &                  &     &   David, + Darryll J(p)      \nl
HASEGAWA        &     T   &  11.27 & 286 & 3224 & C &                  &   Ja&   Takshuito, + others        \nl    
ITOH            &     N   &  13.28 & 242 & 3214 & C &                  &   Ja&   Naoki, Noboru, Nobunari    \nl    
YAMAZAKI        &     T   &   9.60 & 328 & 3148 & C &                  &   Ja&   Takashi, Takaski, Toshitsugu, + others \nl
KUBO            &     Y   &  18.75 & 160 & 3000 & C &                  &   Ja&   Yoshio, + others                          
\enddata
\end{deluxetable}